\documentclass[11pt, a4paper]{article}
\usepackage{styleBuding}
\usepackage{amsmath}
\usepackage{bm}
\usepackage{amssymb}
\usepackage{listings}
\usepackage{multirow}
\usepackage{makecell}
\usepackage{array}
\usepackage{subfigure}
\usepackage{siunitx} 
\usepackage{booktabs}

\usepackage{csquotes}

\usepackage[T1]{fontenc} 
\usepackage{slashed}
\def\be{\begin{equation}}
\def\ee{\end{equation}}
\def\bea{\begin{array}}
\def\eea{\end{array}}
\def\beqa{\begin{eqnarray}}
\def\eeqa{\end{eqnarray}}
\def\beqas{\begin{eqnarray*}}
\def\eeqas{\end{eqnarray*}}

\def\bp{\begin{picture}}
\def\ep{\end{picture}}
\def\bc{\begin{center}}
\def\ec{\end{center}}
\def\bfig{\begin{figure}}
\def\efig{\end{figure}}

\def\bit{\begin{itemize}}
\def\eit{\end{itemize}}

\def\[{\left[}
\def\]{\right]}
\def\({\left(}
\def\){\right)}

\def\..{\left.}
\def\.{\right.}

\def\ep{\epsilon}

\allowdisplaybreaks[1]

\title{\boldmath Light higgsino scenario confronted with muon g-2}

\author[a]{Jun Zhao}
\author[a]{,~Jingya Zhu}
\author[b]{,~Pengxuan Zhu}
\author[b,c]{,~Rui Zhu}

\affiliation[a]{Joint Research Center for Theoretical Physics, School of Physics and Electronics, 
Henan University, Kaifeng 475004,  P. R. China}
\affiliation[b]{CAS Key Laboratory of Theoretical Physics, Institute of Theoretical Physics, 
Chinese Academy of Sciences, Beijing 100190, P. R. China}
\affiliation[c]{School of Physics Sciences, University of Chinese Academy of Sciences,  
Beijing 100049, P. R.China}

\emailAdd{junzhao@henu.edu.cn}
\emailAdd{zhujy@henu.edu.cn}
\emailAdd{zhupx99@icloud.com}
\emailAdd{zhurui@itp.ac.cn}

\abstract{ Light higgsinos below several hundred GeV are favored or required by the 
naturalness of low energy supersymmetry. If only higgsinos are light while other sparticles are sufficiently heavy, 
we have the so-called light higgsino scenario. Confronted with the muon $g-2$ data, this scenario is examined in 
this work. Since in this scenario the LSP (lightest sparticle) is higgsino-like, we need to also consider the 
dark matter constraints. Assuming a light higgsino mass parameter $\mu$ in the range of 100-400 GeV while gaugino
mass parameters above TeV, we explore the parameter space under the muon $g-2$ data and the dark matter constraints. 
We find that, to explain the muon $g-2$ anomaly at $2\sigma$, the winos and sleptons are respectively upper bounded 
by 3 TeV and 800 GeV. In this case, we find that the light higgsino-like dark matter can sizably scatter with 
nucleon and thus the allowed parameter space can be covered almost fully by the future LZ dark matter detection 
project. We also perform a Monte Carlo simulation to figure out the potential of HL-LHC to detect the light sleptons 
in this scenario. It turns out that compared with the current LHC limits, the HL-LHC can further cover a part 
of the parameter space. 
}

\begin{document} 
\maketitle
\flushbottom

\section{Introduction}
\label{sec:intro}
The lastest Fermilab result of muon $g-2$~\cite{FNAL:gmuon}, combined with the BNL result ~\cite{BNL:gmuon}, is 
4.2$\sigma$ above the SM prediction~\cite{Aoyama:2020ynm,RBC:2018dos}, i.e., the experimental value exceeds the
SM value by $(2.51\pm 0.59) \times 10^{-9}$. For new physics researchers, this anomaly is just 
like that a long drought meets a shower of rain. As said by Steven Weinberg forty years ago, 
{\it `Physics thrives on crisis. We all recall the great progress made while finding a way out of the 
various crises of the past'}~\cite{Weinberg:1988cp}. Now to move on beyond the Standard Model (SM) we indeed need 
crisis from experiments. Such an anomaly from the muon $g-2$ measurement may serve as a crisis (albeit some lattice 
calculations \cite{Colangelo:2022vok,Ce:2022kxy,Borsanyi:2020mff,FermilabLattice:2022izv,Alexandrou:2022amy} 
seem to shift up the SM value to relax this crisis, which, however, may transfer 
the crisis to the electroweak fit \cite{Passera:2008jk,Crivellin:2020zul,Colangelo:2020lcg}) and suggest the direction of new physics. 
Not surprisingly, this muon $g-2$ anomaly caused a sensation in high energy physics and  
attempts of explanation have been made in various new physics theories. 

As a leading new physics candidate, the low energy supersymmetry (SUSY) has been recently revisited  
(for recent brief reviews, see, e.g., \cite{Wang:2022rfd,Baer:2020kwz,Yang:2022qyz}) to explain the 
muon $g-2$ anomaly, albeit the SUSY contribution was calculated long ago~\cite{moroi,Stockinger:2006zn, Martin:2001st}. 
(i) In the low energy effective SUSY models, such as the minimal supersymmetric model (MSSM) \cite{Haber:1984rc, Cao:2011sn} 
and the next-to-minimal supersymmetric model (NMSSM) \cite{Ellwanger:2009dp,Cao:2012fz, Cao:2011sn}, the explanation of the muon $g-2$ anomaly can be readily achieved \cite{Chakraborti:2020vjp, Chakraborti:2021kkr, Chakraborti:2021dli, Chakraborti:2021squ, Chakraborti:2021ynm, Chakraborti:2021mbr, Chakraborti:2022sbj, Abdughani:2019wai, Cox:2018qyi, Athron:2021iuf, Wang:2021bcx, Ning:2017dng, Abdughani:2021pdc, Cao:2021tuh, Wang:2021lwi, Cao:2022htd, Tang:2022pxh, Cao:2018rix, Cao:2022chy, Cao:2022ovk}.
Also, the explanation can be made in other extensions 
of the MSSM \cite{Zhao:2021eaa,Yang:2021duj,Zhang:2021gun,Cao:2019evo,Cao:2021lmj,Wang:2022wdy,Li:2021poy,Barman:2022jdg}.           
(ii) The SUSY models with boundary conditions at some high energy scale, such as the CMSSM or mSUGRA, cannot explain the muon $g-2$ anomaly because of the correlation between the masses of the sparticles \cite{Wang:2021bcx,Chakraborti:2021bmv,Aboubrahim:2021phn}. In order to accommodate the muon $g-2$, extensions are needed, such as extending mSUGRA to gluino-SUGRA \cite{Zhu:2016ncq,Akula:2013ioa,Li:2021pnt,Wang:2015rli,Wang:2018vrr,Ahmed:2021htr}, extending GMSB by introducing Higgs-messenger coupling \cite{Kang:2012ra,Evans:2012hg} or deflecting AMSB \cite{Wang:2015nra,Wang:2016otm}, the flipped $SU(5)$ intersecting D-branes model~\cite{Lamborn:2021snt}. 

Note that in these SUSY explanations of the muon $g-2$, the lightest sparticle (LSP) as the dark matter candidate 
is usually assumed to be bino-like due to the limits from the relic density and direct detections of dark matter.
Very recently, the wino-higgsino admixture, with both winos and higgsinos below several hundred GeV, was 
considered as the dark matter candidate to explain the muon $g-2$ anomaly \cite{Iwamoto:2021aaf}, where it was found that 
the favored parameter space in the spontaneously broken SUGRA is detectable in the future dark matter 
direct detections or the LHC searches for the electroweakinos. 
In this work, we will examine another fascinating scenario, the so-called light higgsino scenario, for the explanation 
of the muon $g-2$ anomaly. In this scenario only higgsinos are light, with the higgsino mass parameter $\mu$ being
usually in the range of 100-300 GeV, favored or required by the naturalness of SUSY 
(see, e.g., \cite{Tata:2020afe}). This scenario provides a higgsino-like dark matter (since gauginos are above TeV, 
significantly heavier than higgsinos), differing from the scenario with wino-higgsino admixture as the dark matter 
considered in \cite{Iwamoto:2021aaf}.    
In our study, assuming a light higgsino mass parameter $\mu$ in the range of 100-400 GeV while gaugino
mass parameters above TeV, we will explore the parameter space by considering the muon $g-2$ data and the 
dark matter constraints. Then for the favored parameter space we will check the detectability of future 
dark matter detection projects and the HL-LHC.

This work is organized as follows. In Sec.~\ref{sec:scenario} we provide a brief description for the 
light higgsino scenario. 
In Sec.~\ref{sec:scan} we perform a numerical scan to locate the parameter space for the explanation of
the muon $g-2$ anomaly at $2\sigma$ level.   
In Sec.~\ref{sec:LHC}  we perform a Monte Carlo simulation to show the HL-LHC detectability for 
the parameter space favored by the muon $g-2$.
Finally, we conclude in Sec.~\ref{sec:conclusions}.

\section{\label{sec:scenario}A brief description of light higgsino scenario} 

From the naturalness of SUSY, the higgsino mass parameter $\mu$ cannot be heavy, which can be seen
from the following minimization relation of the tree-level Higgs potential \cite{Arnowitt:1992qp}
\begin{eqnarray}
\frac{m^2_{Z}}{2}=-\mu^{2}+\frac{M^2_{H_d}-M^2_{H_u}\tan^{2}\beta}{\tan^{2}\beta-1},
\label{minimization}
\end{eqnarray}
with $M^2_{H_d}$ and $M^2_{H_u}$ being the soft SUSY breaking masses of
the Higgs fields at weak scale, while $\mu$ being the mass parameter of the higgsinos. 
Obviously, the value of $\mu$ cannot be too large compared with the weak scale 
in order to avoid fine-tuning, which in natural SUSY \cite{Tata:2020afe} is assumed to be smaller than 300 GeV. Since naturalness has no strict criterion, we in this work assume
$\mu$ in the range of $100-400$ GeV.   
 
In the light higgsino scenario, the gauginos are heavy (above TeV) 
and thus the LSP as the dark matter candidate is utterly dominated by higgsinos.
Not only the LSP is higgsino-like, but the next-to-lightest sparticles ( the neutralino $\tilde{\chi}^0_2$ 
and chargino  $\tilde{\chi}^{\pm}_1$) are also higgsino-like, all of which are nearly degenerate, having a mass
around the value of $\mu$. These higgsino-dominated electroweakinos are approximately given as follows~\cite{Cao:2021lmj}:
\begin{equation}
     m_{\tilde{\chi}_{1}^\pm} \sim \mu, \quad m_{\tilde{\chi}_{1, 2}^0} = \mu \mp \Delta m, \quad \Delta m \simeq \frac{g_1^2 v^2 M_1}{M_1^2 - \mu^2} + \frac{g_2^2 v^2 M_2}{M_2^2 - \mu^2},
\end{equation}
with $v=246~{\rm GeV}$ being the Higgs vacuum expectation value, $M_1$ and $M_2$ being the soft SUSY breaking masses of bino and wino fields, respectively. When $M_1$ and $M_2$ are both greater than $1~{\rm TeV}$, the mass splitting $\Delta m$ is less than $1~{\rm GeV}$. In this case, since the visible particles from the higgsino decay are too soft, the productions of these higgsino-like sparticles at the colliders merely give missing energy and can only be detected by requiring one initial state radiation (ISR) jet, e.g., at the LHC the signal of monojet plus missing energy \cite{Han:2013usa}.
Since such productions of higgsino-like electroweakinos are proceeded by electroweak interaction,
a global likelihood analysis showed that no clear range of their masses can be robustly excluded 
by current LHC searches \cite{GAMBIT:2018gjo}.  Very recently, the LHC constraints on the electroweakinos 
were revisited \cite{Buanes:2022wgm}, which showed that these higgsino-like electroweakinos as light as 
100 GeV are still allowed.
 
Since in this scenario gauginos are heavy (above TeV), the LSP as a dark matter candidate is dominated by
light higgsinos. Such light higgsino-like LSPs can efficiently annihilate, e.g., through the $s$-channel 
$Z$-boson exchange,  to have a large annihilation rate in the early universe.
Thus they usually give a thermal relic density under abundance.  
This implies that these light higgsino-like 
LSPs are only a component of dark matter while other components like axions are needed.    
Therefore, in this scenario the LSP-nucleon scattering cross section must be re-scaled by a factor
$\Omega_{\rm LSP}h^2/\Omega_{\rm PL}h^2$, with $\Omega_{\rm PL}h^2$ being the observed relic
density by Planck satellite.

\section{\label{sec:scan}Parameter space for muon $g-2$}
In our scan, we assume light higgsinos and heavy gauginos 
\begin{eqnarray}
100{\rm ~GeV} \leq \mu \leq 400{\rm ~GeV},~~ 1{\rm ~TeV} \leq M_1 ,M_2 \leq 5{\rm ~TeV}.
\end{eqnarray}
Since the muon $g-2$ is also sensitive to slepton masses, we scan over it from 200 GeV to 2 TeV 
\begin{eqnarray}
 200{\rm ~GeV} \leq M_{L_{\ell}}=M_{E_{\ell}} \leq 2 {\rm ~TeV},
\end{eqnarray}
where $\ell=e,\mu$. 
For the third-generation squark mass parameters, we require them to be heavy due to the 125 GeV Higgs boson mass 
\begin{eqnarray}
3{\rm ~TeV} \leq M_{Q3}, M_{U3} \leq 5{\rm ~TeV} .
\end{eqnarray}
For the third-generation sfermion trilinear couplings, we require them in the range
\begin{eqnarray}
-3 {\rm ~TeV} \leq A_{t,b,\tau} \leq 3 {\rm ~TeV} .
\end{eqnarray}
For the value of $ \tan\beta$, we scan over it in the range of 
$1 \leq \tan\beta \leq 50$.
Other soft mass parameters are fixed at 5 TeV except for the trilinear ones like $A_{u,d,e}$ which are set to zero. 
In our scan we consider the following experimental constraints:
\begin{itemize}

\item[(1)] The package \textsc{SUSY-HIT}~\cite{Djouadi:2006bz} is used for generating the particle spectrum,  where the mass spectrum is calculated by subprogram \texttt{SuSpect-2.41}, and the decays of the Higgs boson and sparticles are calculated by subprogram \texttt{HDECAY-3.4} and \texttt{SDECAY-1.5}, respectively. The Higgs boson masses are evaluated with two-loop corrections, under the approximations of vanishing external momenta and of vanishing EW gauge couplings. So the SM-like Higgs boson mass is required in the range of $122 < m_h < 128$ {\rm ~GeV} \footnote{The accuracy of Higgs mass calculation has been improved to state-of-the-art. Now the theoretical uncertainties are understandable, and in most recent studies, the total Higgs mass uncertainty was improved to less than 2 GeV for low-energy MSSM parameter space. For the relevant studies, see, e.g., ~\cite{Degrassi:2002fi, Bahl:2019hmm, Bagnaschi:2017xid, Athron:2016fuq, Allanach:2004rh, Gogoladze:2011aa, AdeelAjaib:2013dnf, Drechsel:2016htw}. In this work, we adopted 3 GeV as the accuracy in the code \textsc{SuSpect}.}.

\item[(2)] We consider the constraint of meta-stability of the vacuum state, which requires 
$|A_t|\lesssim2.67(m_{\tilde{t}_L}^2+m_{\tilde{t}_R}^2+\mu^2+m_{H_u}^2)$~\cite{vacuum}.

\item[(3)] The sleptons are required to be above 200 GeV, considering the LEP2 plus LHC constraints.

\item[(4)] The LSP dark matter relic density is calculated by \textsc{MicrOMEGAs-5.2.13}~\cite{Belanger:2010gh} 
and is required below its $2\sigma$ upper bound of the Planck observed value 
$\Omega_{\rm DM}h^2=0.120\pm0.001$~\cite{Planck:2018vyg}.

\item[(5)] The two-loop level SUSY contributions to muon $g-2$ are calculated by the package \textsc{GM2Calc-2.1.0}~\cite{Athron:2015rva, Athron:2022gga}. We require SUSY to explain the current data $\Delta a_{\mu}\equiv a_{\mu}^{\rm exp}- a_{\mu}^{\rm SM}=(2.51\pm0.59)\times10^{-9}$~\cite{Muong-2:2021ojo} within the $2\sigma$ range.                  	
\end{itemize}

\begin{figure*}[tbp]
    \centering 
	\includegraphics[width=.48\textwidth]{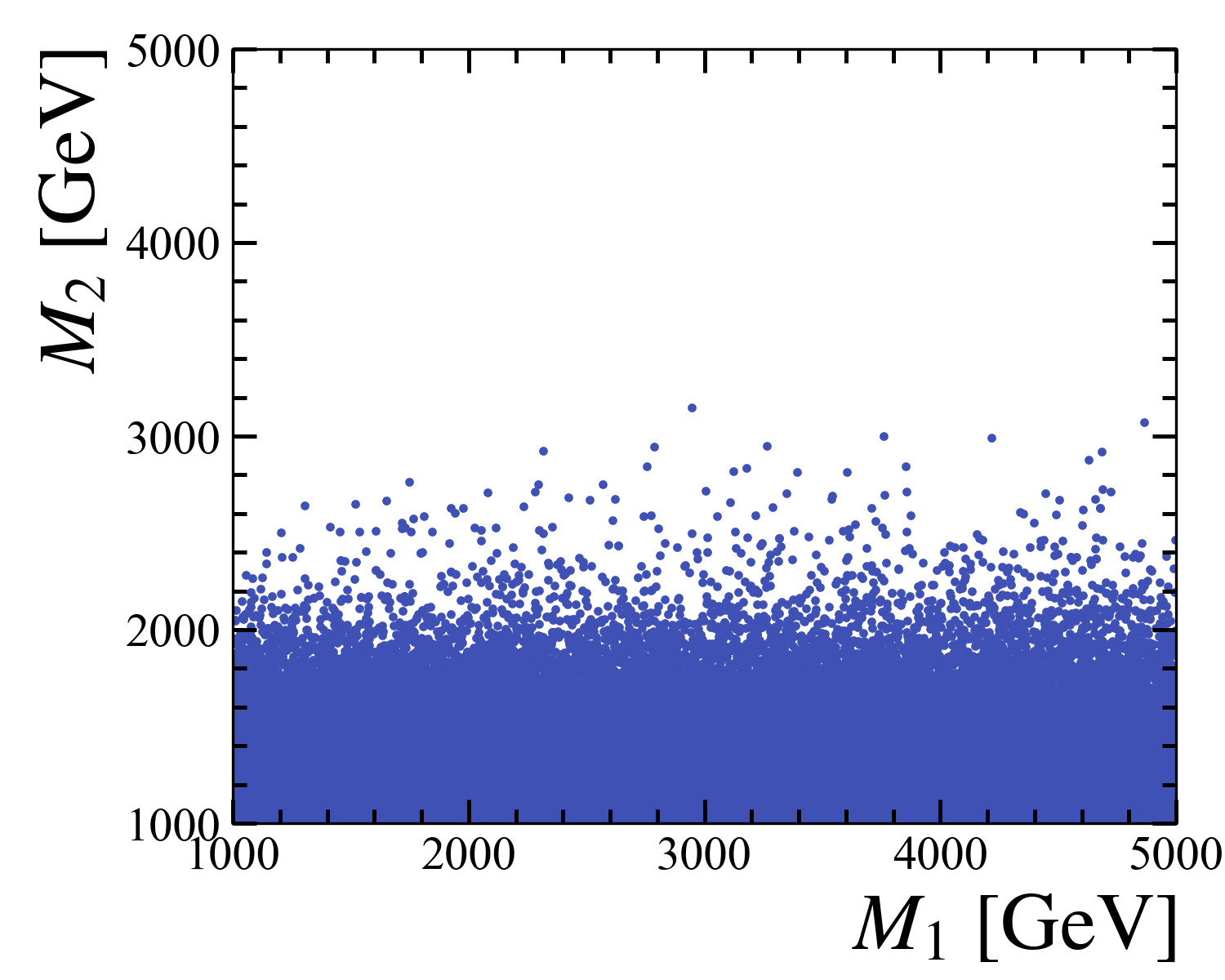}\hspace{0.015\textwidth}
	\includegraphics[width=.48\textwidth]{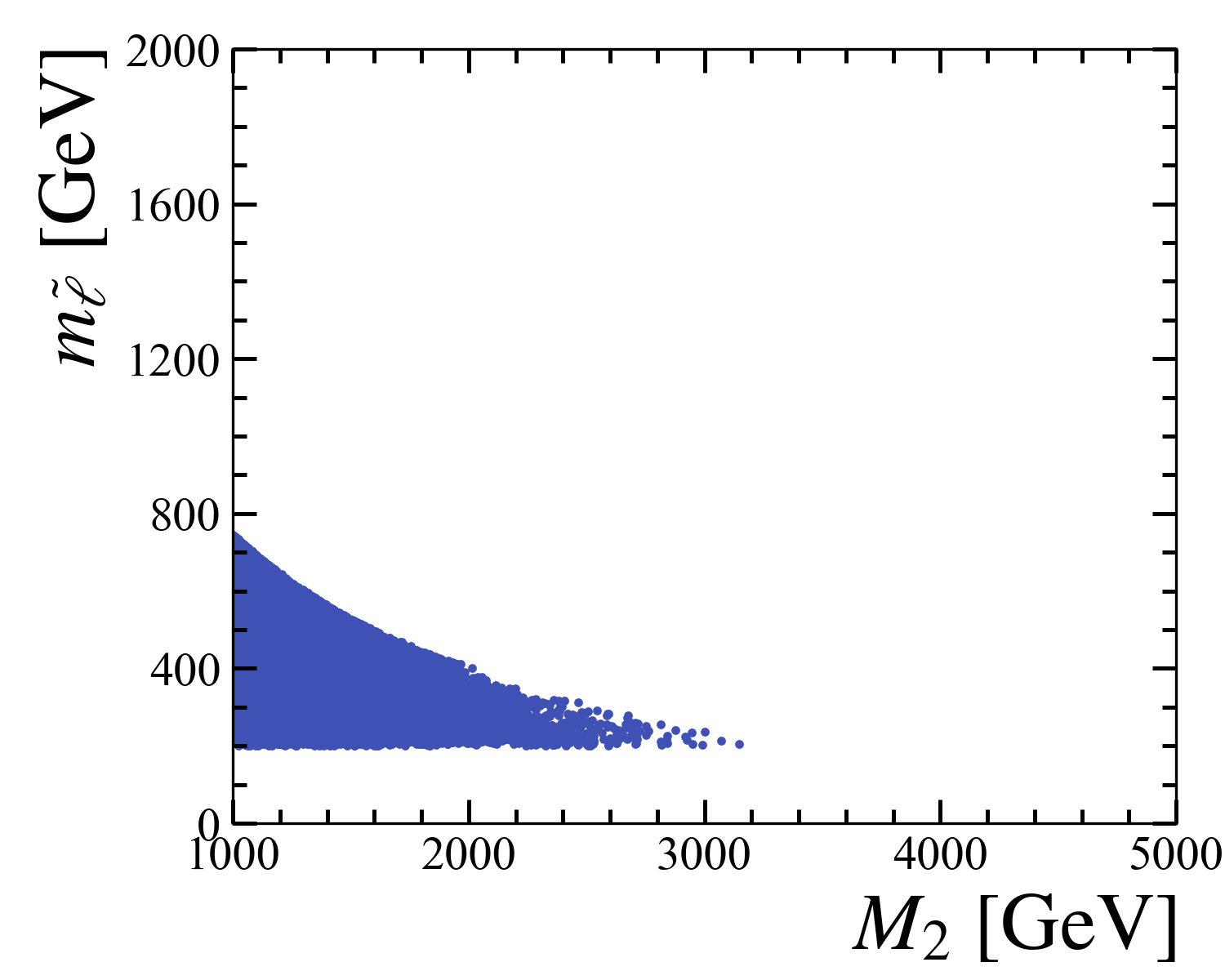}
	\includegraphics[width=.48\textwidth]{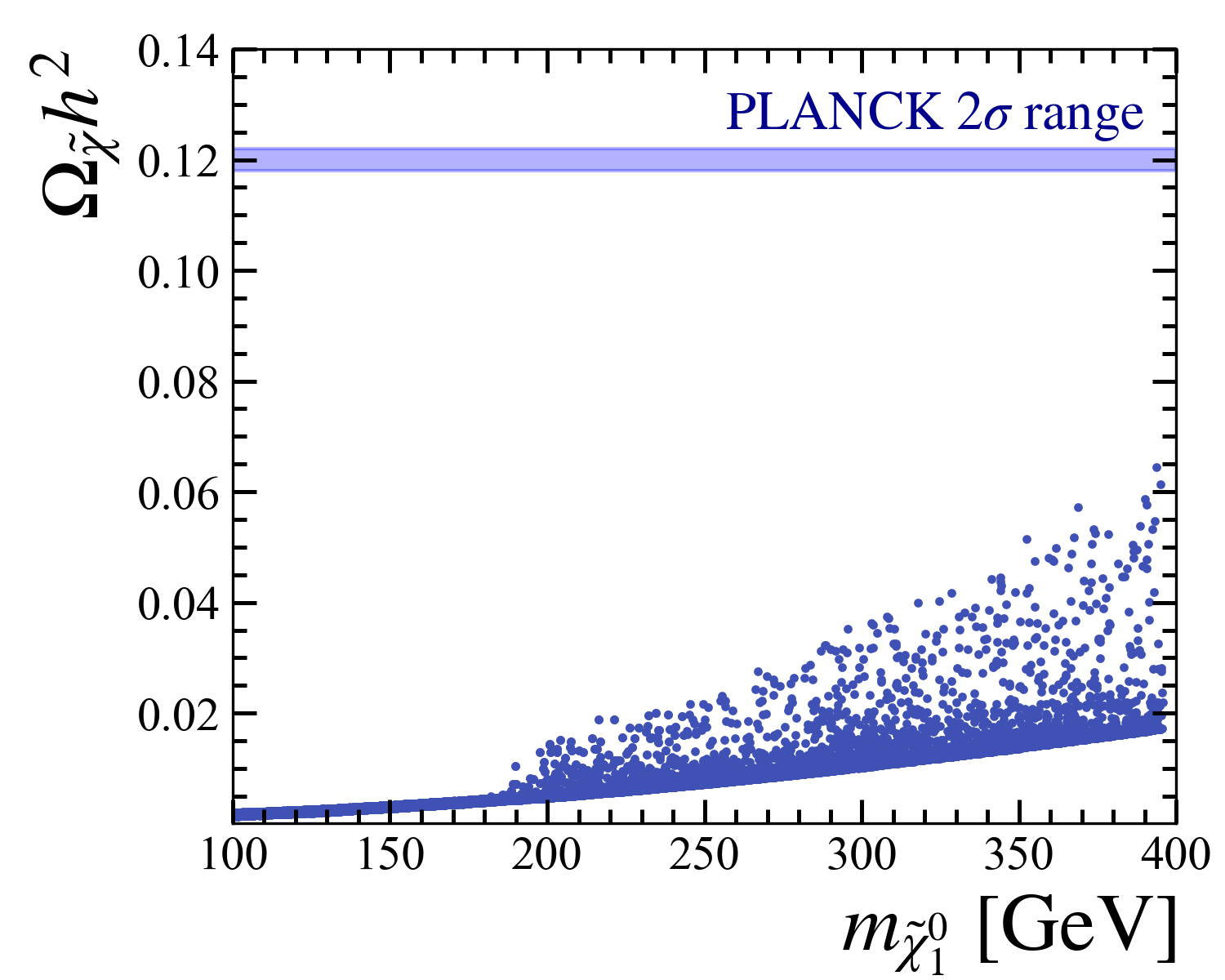}\hspace{0.015\textwidth}
	\includegraphics[width=.48\textwidth]{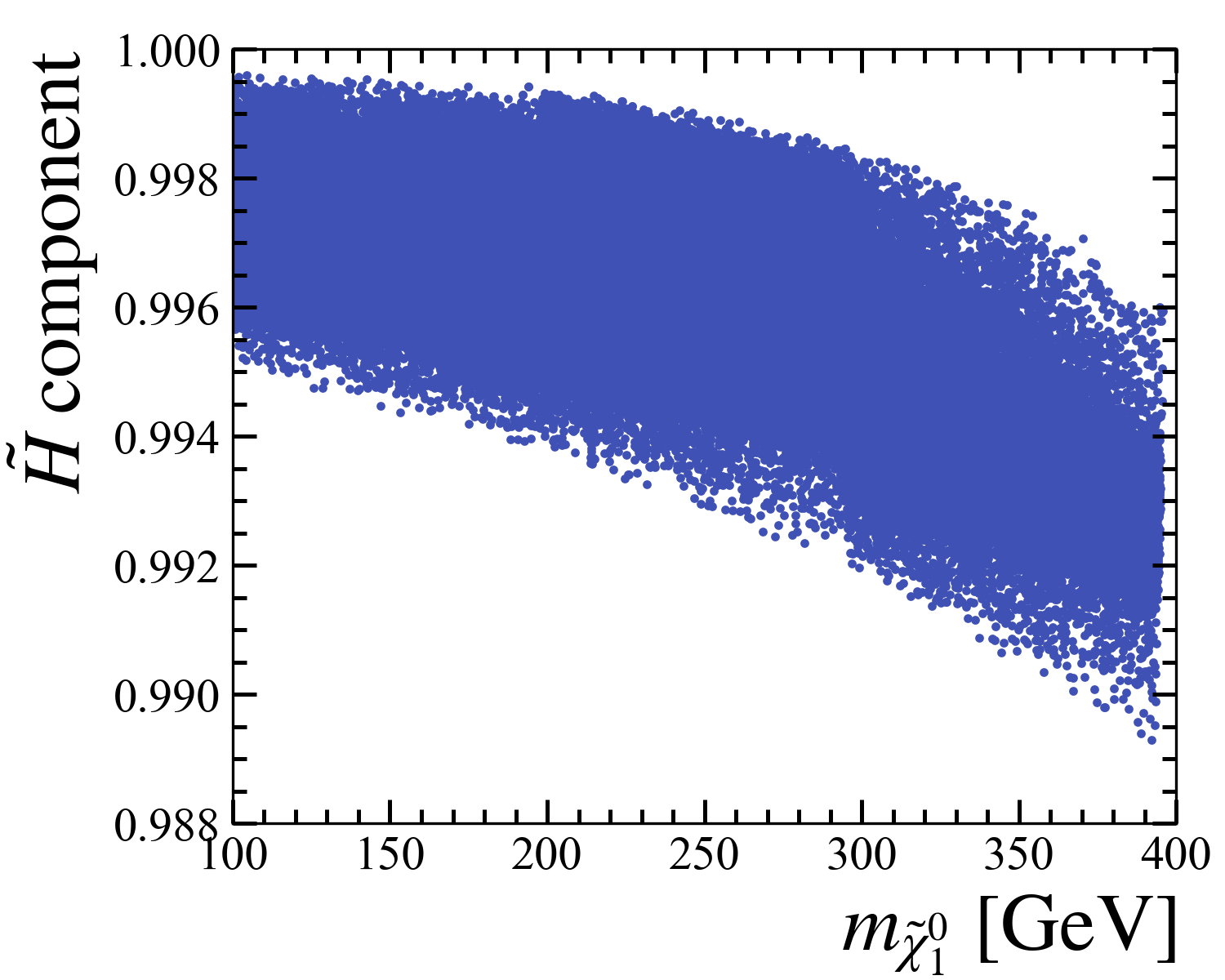}
    \caption{\label{fig:mass1} 
    Scatter plots of the samples survived the constraints (1-5) including especially the muon $g-2$ 
    at $2\sigma$ level.}
\end{figure*}

In Fig.~\ref{fig:mass1} we plot the scatter plots of the samples survived the constraints (1-5) 
including especially the muon $g-2$ at $2\sigma$ level. This figure shows the following characteristics:
\begin{itemize}
\item[(i)] From the top panels we see that in this light higgsino scenario the muon $g-2$ data at $2\sigma$ level requires the wino mass $M_2$ below 3 TeV and the slepton mass below 800 GeV, while it is not 
sensitive to the bino mass $M_1$. 
\item[(ii)] From the bottom-left panel we see that such higgsino-like LSP gives a thermal relic density 
much below the measured abundance.  
\item[(iii)] From the bottom-right panel we see the LSP is utterly dominated by the higgsino component.
\end{itemize} 

\begin{figure}[h]
	\centering
	\includegraphics[width=.6\textwidth]{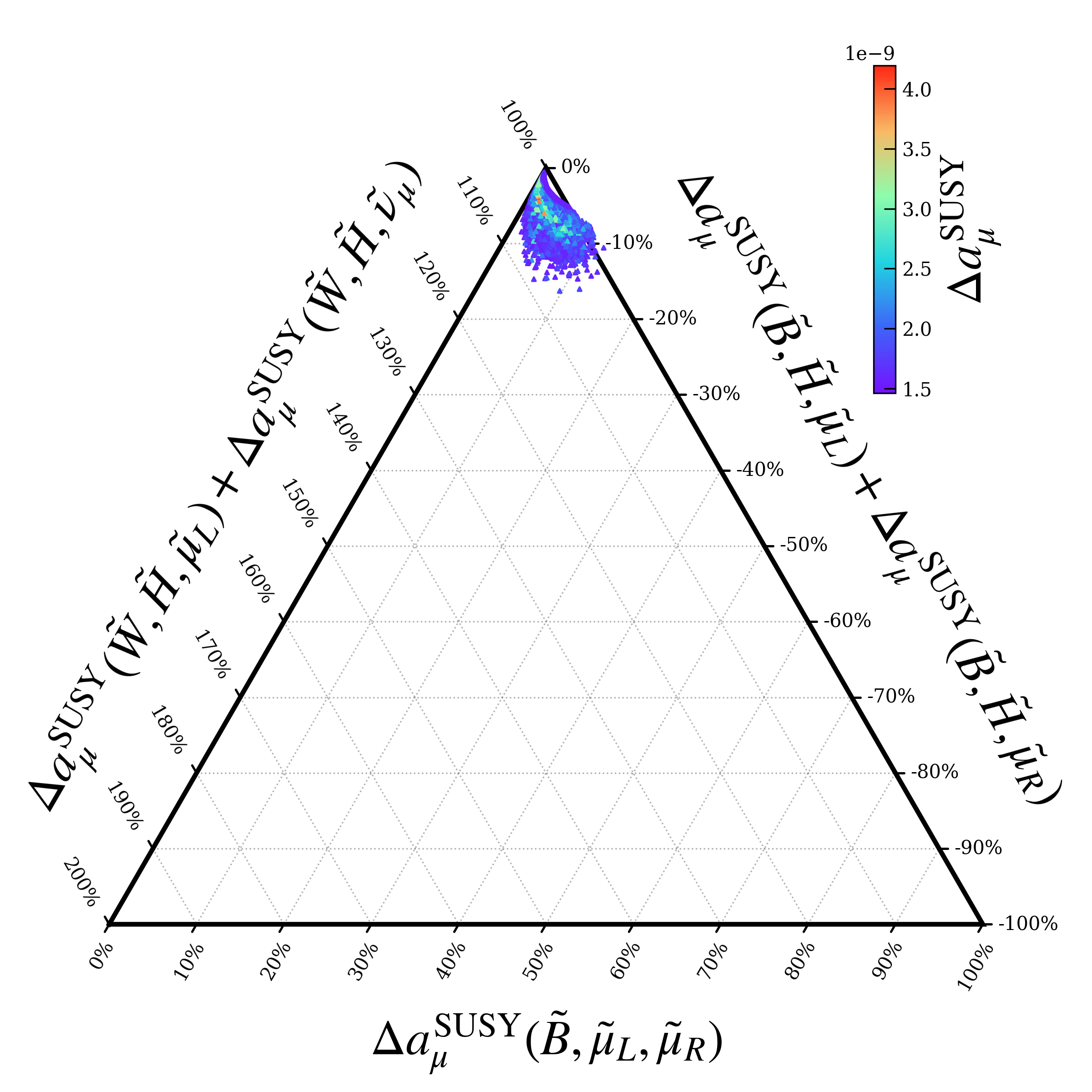}
	\caption{\label{fig:tn_amususy} A ternary plot showing the components of the one-loop SUSY contributions to muon $g-2$. The left axis represents the wino-higgsino loop, the right axis represents the bino-higgsino loops, and the bottom axis represents the bino-smuon loop. The colors are coded by the two-loop result of $\Delta a_\mu^{\rm SUSY}$. }
\end{figure}

\begin{figure*}[tbp] 
    \centering 
    \includegraphics[width=.34\linewidth]{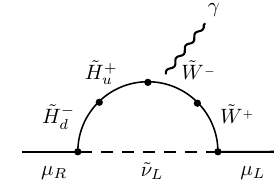}\hspace{.07\linewidth}
    \includegraphics[width=.34\linewidth]{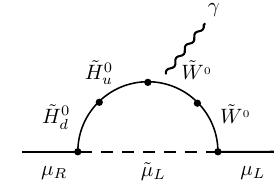}
\vspace{ 0.0cm} 
    \caption{\label{wino-higgsino} 
The muon $g-2$ Feynman diagrams showing the contributions of light higgsinos assisted by not-too-heavy winos.
Here the charged-current loop dominates the contributions.} 
\end{figure*}

The relevant SUSY parameters result in different contributions to $a_\mu^{\rm SUSY}$. The two-loop expressions of $a_\mu^{\rm SUSY}$ implemented in the package \textsc{GM2Calc} take the form in Refs.~\cite{vonWeitershausen:2010zr, Fargnoli:2013zia, Bach:2015doa}. To figure out the dominant SUSY contribution, the detailed five contributions at one-loop level are shown in Fig.~\ref{fig:tn_amususy} via a ternary scatter plot\footnote{A ternary plot depicts the ratios of the three variables as positions in an equilateral triangle. Note that any one of the variables is not independent to the others.} with the colors coded by $\Delta a_\mu^{\rm SUSY}$. As shown in Fig.~\ref{fig:tn_amususy}, the five contributions are classified into three classes: the wino-Higgsino loops $\Delta a_\mu^{\rm SUSY}(\tilde{W},\tilde{H},\tilde{\mu}_L) + \Delta a_\mu^{\rm SUSY}(\tilde{W},\tilde{H},\tilde{\nu}_\mu)$ (left axis), the bino-Higgsino loops $\Delta a_\mu^{\rm SUSY}(\tilde{B},\tilde{H},\tilde{\mu}_L)  + \Delta a_\mu^{\rm SUSY}(\tilde{B},\tilde{H},\tilde{\mu}_R) $ (right axis) and the bino-smuon loops $\Delta a_\mu^{\rm SUSY}(\tilde{B},\tilde{\mu}_L,\tilde{\mu}_R)$ (bottom axis). From Fig.~\ref{fig:tn_amususy}, one can find the dominant SUSY contribution is from wino-higgsino loop. The bino-higgsino loop provides a $\sim 10\%$ negative contribution while the bino-smuon loop provides a $\sim 10\%$ positive contribution. 

Therefore, we can understand why a sizable contribution to the muon $g-2$ in this light higgsino scenario needs a not-too-heavy wino mass. If gauginos are too heavy and decoupled, the light electroweakinos 
will be only higgsinos. The interactions of a higgsino with muon and slepton (smuon or sneutrino) 
are kind of Yukawa couplings, which always flip the chirality between muon and slepton. Hence, 
a pure higgsino cannot flip the muon chirality via loops, except with a mass insertion of left-right handed smuon transition $m_{\mu}\left( A_{\ell} -\mu \tan{\beta} \right)$, 
 which can be neglected. As a result, the pure higgsinos 
cannot make enough contributions to explain the  muon $g-2$ at $2\sigma$ level.  
When not-too-heavy winos come into play, we have Fig.\ref{wino-higgsino}. In this figure, the wino-higgsino loop contributions are given as follows \cite{Athron:2015rva, Martin:2001st, moroi, Stockinger:2006zn}
    \begin{equation}\label{eq:amu-wino-higgsinos}\begin{split}
  \Delta a_\mu^{\rm SUSY}(\tilde{W},\tilde{H},\tilde{\nu}_\mu)
  &\simeq \frac{g_2^2 m_{\mu}^2}{8\pi^2} \frac{M_2 \mu \tan{\beta}}{m_{\tilde{\nu}_{\mu}}^4} \cdot F_a\left(\frac{M_2}{m_{\tilde{\nu}_\mu}}, \frac{\mu}{m_{\tilde{\nu}_\mu}}\right),\\
 \Delta a_\mu^{\rm SUSY}(\tilde{W},\tilde{H},\tilde{\mu}_L) &\simeq -\frac{g_2^2 m_{\mu}^2}{16\pi^2} \frac{M_2 \mu \tan{\beta}}{m_{\tilde{\mu}_{L}}^4} \cdot
 F_b\left(\frac{M_2}{m_{\tilde{\mu}_L}}, \frac{\mu}{m_{\tilde{\mu}_L}}\right),
\end{split}
\end{equation} 
where the loop functions are defined as 
\begin{equation}\begin{split}
	F_a(x, y) = \frac{1}{2} \frac{G_3(x^2) - G_3(y^2)}{x^2 - y^2},  &\quad F_b(x, y) = -\frac{1}{2} \frac{G_4(x^2) - G_4(y^2)}{x^2 - y^2},  \\
	G_3(x) = \frac{3-4x + x^2 + 3\log{x}}{(1-x)^3}, &\quad G_4(x) = \frac{1-x^2 + 2x \log{x}}{(1-x)^3}. 
\end{split}\end{equation}
They satisfy $0 \leq F_{a, b}(x, y) \leq 1$ and are monochromatically increasing for $x$ and $y$, satisfying $F_a(1, 1) = 1/2$ and $F_b(1, 1) = 1/6$ for degenerate sparticle masses.\footnote{$F_a$ and $F_b$ are reduced from the functions in Ref.~\cite{moroi}. } Other contributions come from the bino loops, where the bino-higgsino loops take approximate forms \cite{Athron:2015rva, Martin:2001st, moroi, Stockinger:2006zn}
\begin{equation}\label{eq:amu-bino-higgsinos}\begin{split}
 \Delta a_\mu^{\rm SUSY}(\tilde{B},\tilde{H},\tilde{\mu}_L) 
 &\simeq \frac{g_1^2 m_{\mu}^2}{16\pi^2} \frac{M_1 \mu \tan{\beta}}{m_{\tilde{\mu}_{L}}^4} \cdot F_b\left(\frac{M_1}{m_{\tilde{\mu}_L}}, \frac{\mu}{m_{\tilde{\mu}_L}}\right),\\
 \Delta a_\mu^{\rm SUSY}(\tilde{B},\tilde{H},\tilde{\mu}_R) &\simeq -\frac{g_1^2 m_{\mu}^2}{8\pi^2} \frac{M_1 \mu \tan{\beta}}{m_{\tilde{\mu}_{R}}^4} \cdot F_b\left(\frac{M_1}{m_{\tilde{\mu}_R}}, \frac{\mu}{m_{\tilde{\mu}_R}}\right),
\end{split}
\end{equation} 
From Eq.~(\ref{eq:amu-wino-higgsinos}) and Eq.~(\ref{eq:amu-bino-higgsinos}), one can find that the contributions of the wino-higgsino loops are positive while the contributions of the bino-higgsino loops tend to be negative.
Since bino in this work is heavier than $1~{\rm TeV}$ and its contribution is proportional to $g_1^2$, about a quarter of $g_2^2$, so not-too-heavy bino cannot play a dominated role. 

\begin{figure*}[tbp]
    \centering 
    \includegraphics[width=.7\textwidth]{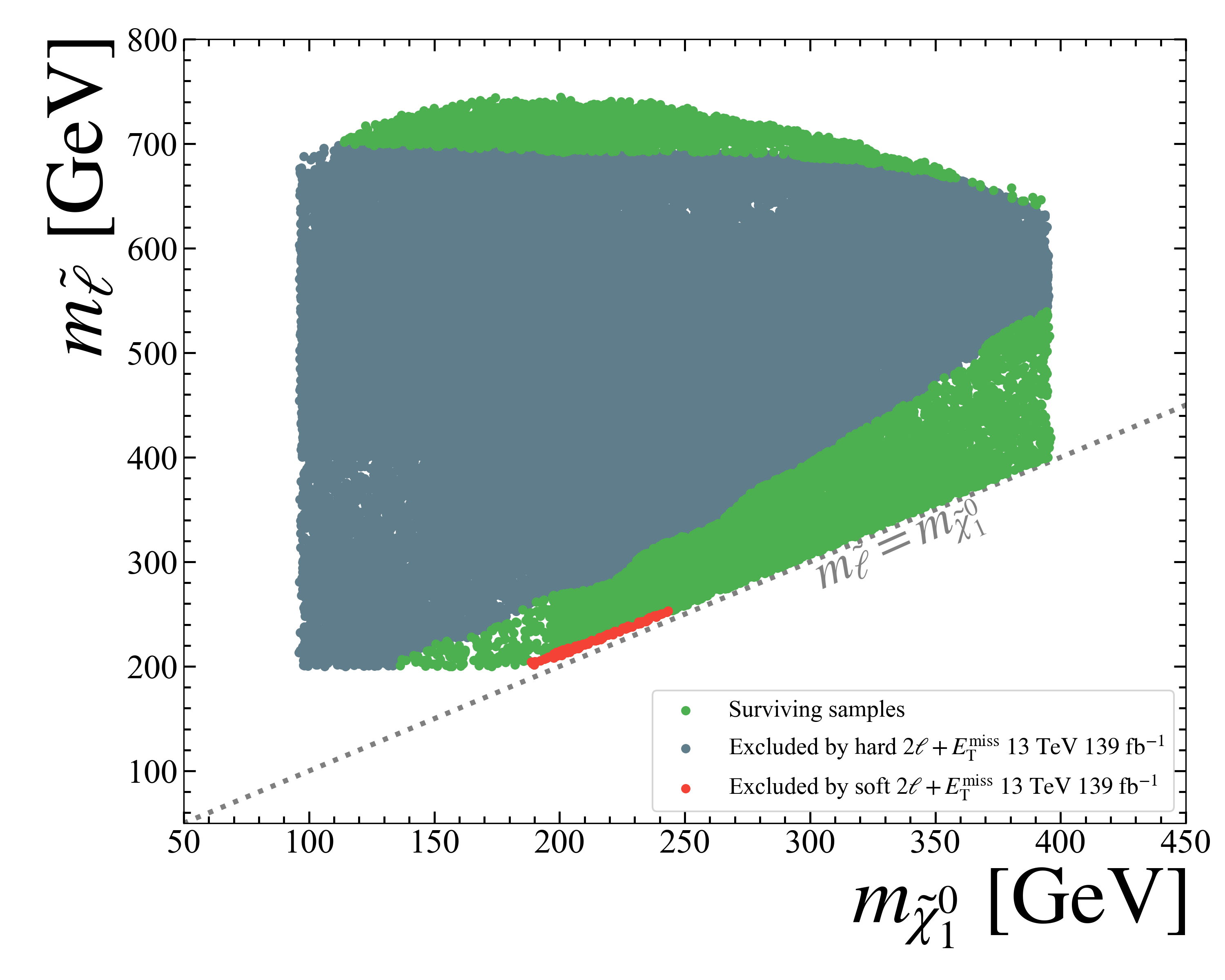}
    \caption{\label{fig:mass2} 
Same as Fig.~\ref{fig:mass1}, but displayed on the plane of slepton mass versus the LSP mass.
The regions excluded by the ATLAS searches for the hard dilepton plus missing energy  
~\cite{ATLAS:2019lff} and soft dilepton plus missing energy~\cite{ATLAS:2019lng} are shown.  }
\end{figure*}

In Fig.~\ref{fig:mass2} we show the samples survived the constraints (1-5)  on the plane of 
slepton mass versus the LSP mass.
The regions excluded by the ATLAS searches for the two hard leptons plus missing energy  
~\cite{ATLAS:2019lff} and two soft leptons plus missing energy~\cite{ATLAS:2019lng} are displayed.
We see that for a large mass splitting between slepton mass and LSP mass, the ATLAS searches for 
the two hard leptons plus missing energy have excluded a quite large part of the parameter
space required for the explanation of  the muon $g-2$ at $2\sigma$ level.    
While for a compressed slepton-LSP spectrum, i.e., a very small mass splitting between 
slepton mass and LSP mass, the exclusion ability of the current LHC is rather limited.

\begin{figure*}[tbp]
    \centering 
    \makebox[\textwidth][c]{
    \includegraphics[width=.33\textwidth]{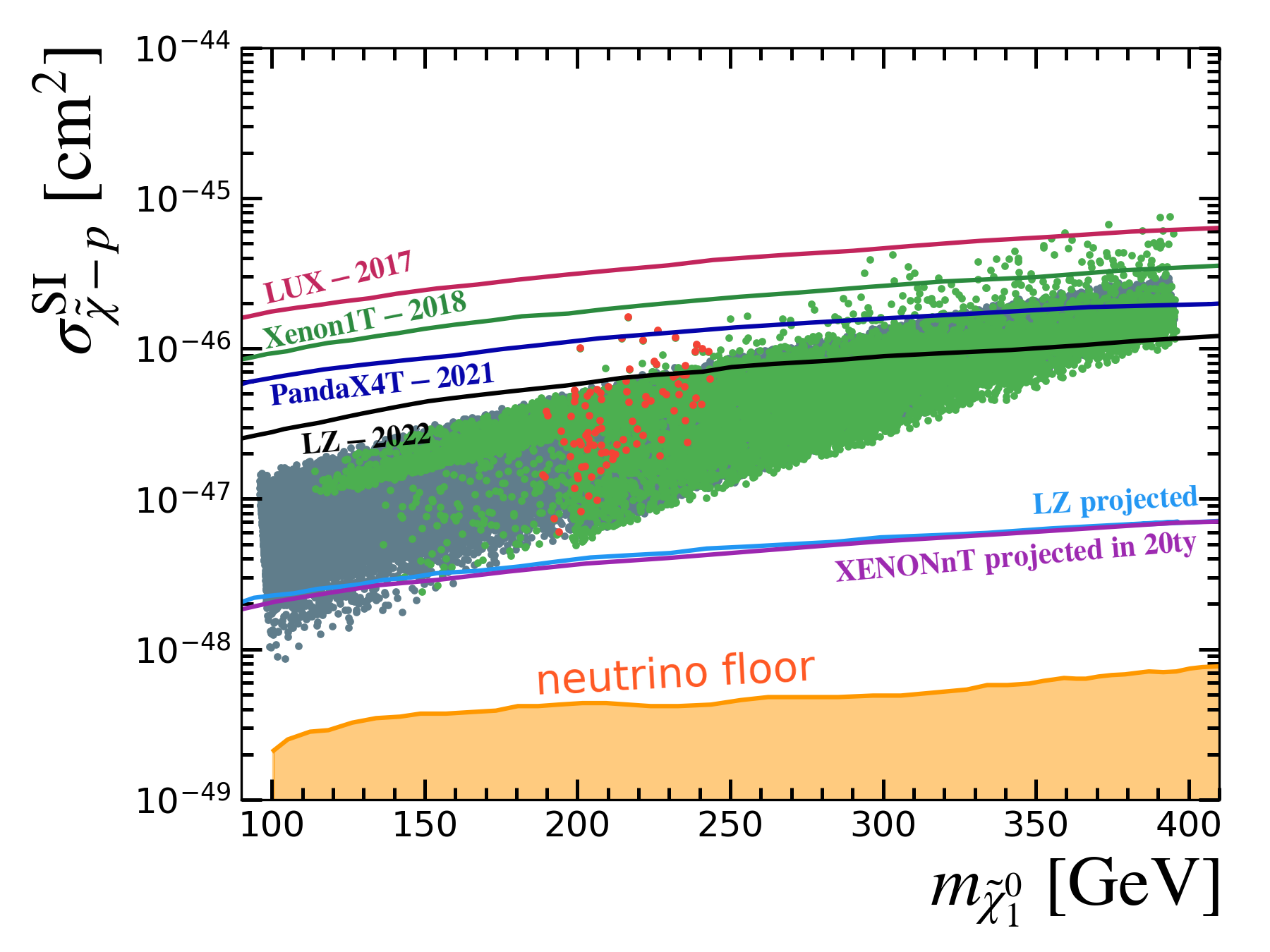}
    \includegraphics[width=.33\textwidth]{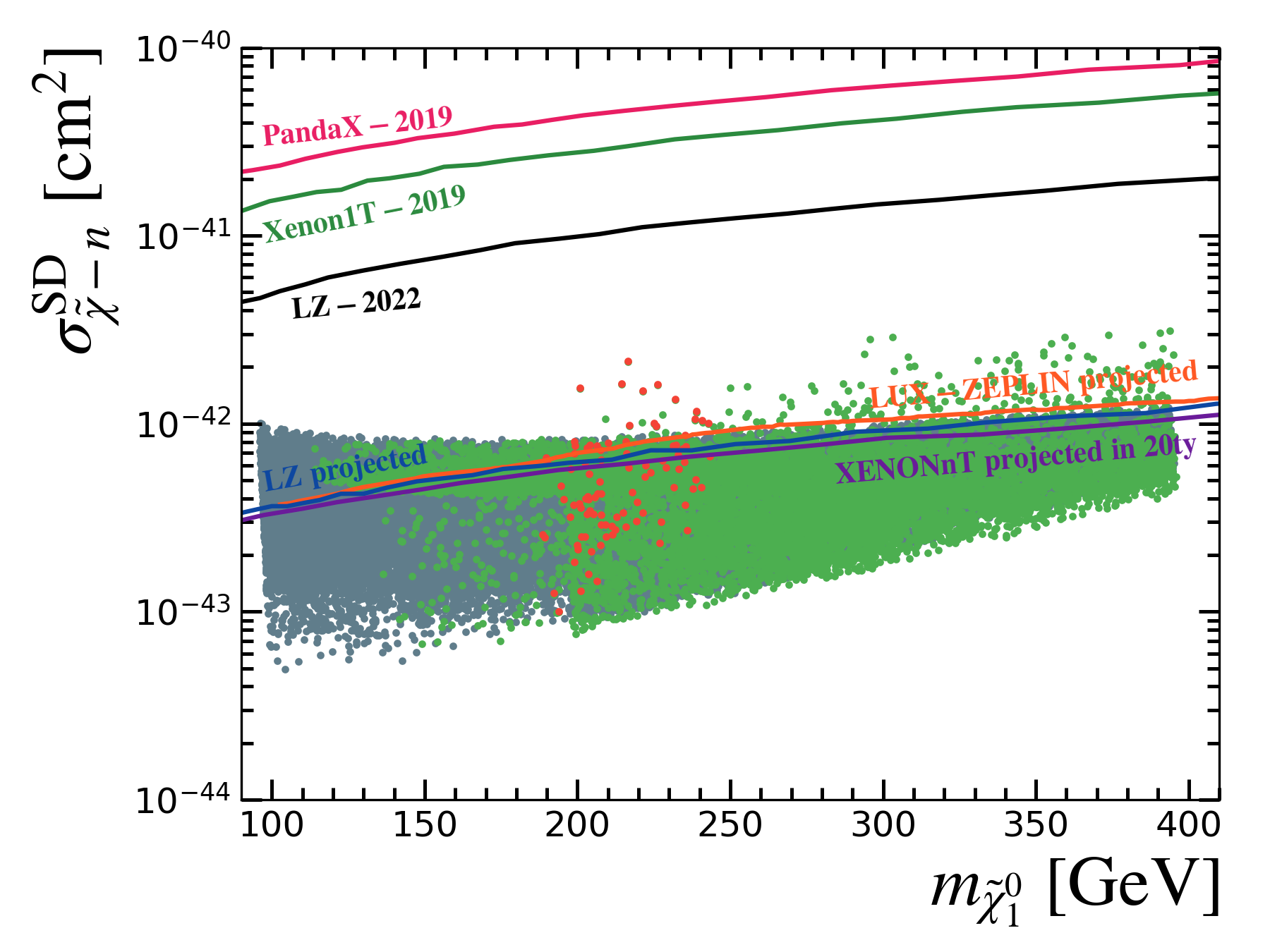}
    \includegraphics[width=.33\textwidth]{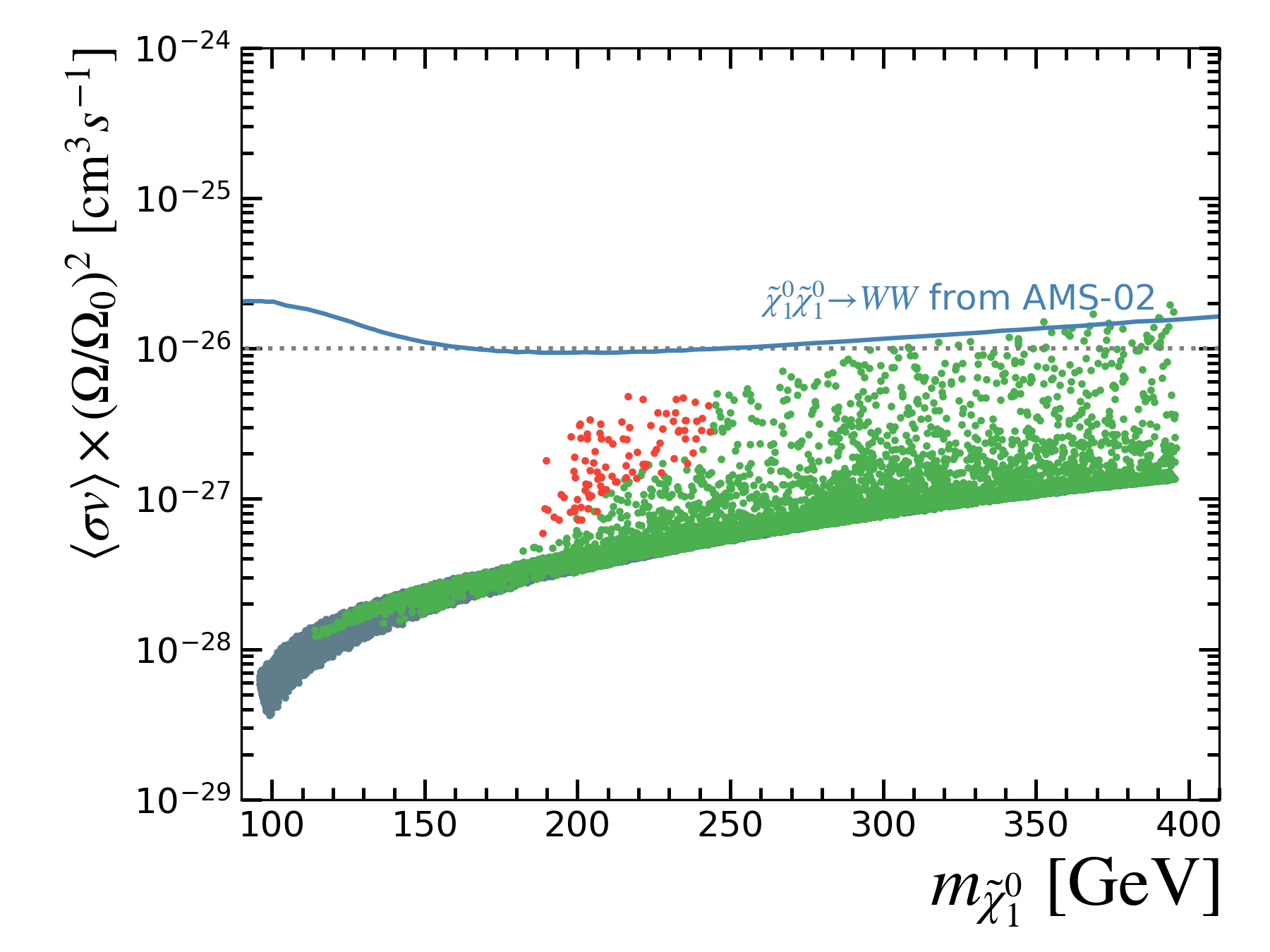}
    }
    \caption{\label{fig:fig-SISD} 
Same as the samples in Fig.~\ref{fig:mass2}, but showing the spin-independent and spin-dependent 
LSP-nucleon scattering cross sections and DM annihilation cross sections versus the LSP mass. Since the higgsino-like LSP is underabundant, 
a scaling factor $\Omega_{\rm LSP}h^2/0.12$ is applied on the LSP-nucleon scattering cross sections.
The $90\%$ C.L. upper limits from LUX-2017~\cite{LUX2017}, XENON1T-2018~\cite{XENON:2018voc},XENON1T-2019~\cite{XENON:2019rxp}, PandaX-2019~\cite{PandaX-II:2018woa}, PandaX4T-2021~\cite{PandaX-4T:2021bab},
LZ-2022~\cite{LZ:2022ufs}
as well as the future sensitivities from  
LZ~\cite{LZ:2018qzl}, XENONnT(20ty)~\cite{XENON:2020kmp} are shown. The $95\%$ C.L. constraints on DM annihilation into $WW$ are derived from the anti-proton and B/C data of AMS-02~\cite{Reinert:2017aga}.} 
\end{figure*}

In Fig.~\ref{fig:fig-SISD}, we replot the samples of Fig.~\ref{fig:mass2}, showing the spin-independent and spin-dependent LSP-nucleon scattering cross sections and the velocity-averaged DM annihilation cross section versus the LSP mass. In the aspect of DM direct detections, the current direct detection limits are shown assuming that a single dark matter candidate constitutes the entire relic.  Since the higgsino-like LSP is under-abundant,  so the direct detection limits are applied on the scaled cross sections, where the scaling factor is $\Omega_{\rm LSP}h^2/0.12$.   We see that after scaling the most samples of the  higgsino-like LSP can survive the current direct detection limits, while the future LZ-projected can almost cover all the survived samples. Most of the DM indirect detection experiments are searching for the DM self-annihilation rate $\Gamma_{A}$ given by 
    \begin{equation}
    	\Gamma_{A} \propto \langle \sigma v \rangle \times \frac{\rho_{\rm DM}^2}{m_{\rm LSP}^2},
    \end{equation}
    where $\rho_{\rm DM}$ is the DM density in the local halo. 
     We checked that higgsino-like LSPs mostly annihilate into gauge bosons (over 80\% in most case) and/or Higgs bosons. These primal annihilation products subsequently decayed and can act as the sources of some cosmic ray flux, such as positrons, anti-protons and photons, and neutrino flux in our galaxy. From Fig.~\ref{fig:fig-SISD}, one can find that the annihilation rate is diluted by the square of the local density scaling factor. The current experiments, e.g. the constraints from AMS-02~\cite{Reinert:2017aga}, are hard to detect $\langle \sigma v \rangle$ below $10^{-26}~{\rm cm}^3 s^{-1}$ for $m_{\tilde{\chi}_1^0}$ larger than 100 GeV. Although the limits in Fig.~\ref{fig:fig-SISD} depend on the properties of the DM halo and the cosmic ray propagation, etc, one can argue that the current DM indirect detection experiments are not sensitive to the light higgsino scenario. 

\section{\label{sec:LHC}Detectability at the HL-LHC}  
\begin{figure*}[tbp]
    \centering 
    \includegraphics[width=.4\textwidth]{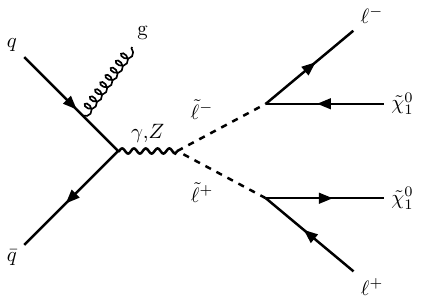}
    \caption{\label{fig:fig-ferman} 
A typical Feynman diagram for the slepton pair production process 
$pp\to j\tilde{\ell}^+ (\to \ell^+ \tilde\chi^0_1) \tilde{\ell}^-(\to \ell^- \tilde\chi^0_1) \to j+\ell^+\ell^- + E_{\rm T}^{\rm miss}$ at the LHC. }
\end{figure*}

Since the muon $g-2$ data requires sleptons below 800 GeV, we examine the observability of the
slepton pair production at the 14 TeV HL-LHC with 3000~${\rm fb}^{-1}$. 
For this end, we perform a detailed Monte Carlo simulation for the process
 \begin{equation} 
 pp\to j \tilde{\ell}^+ (\to \ell^+ \tilde\chi^0_1) \tilde{\ell}^-(\to \ell^- \tilde\chi^0_1) \to j+\ell^+\ell^- + E_{\rm T}^{\rm miss}.
\end{equation} 
A typical Feynman diagram of this process is shown in Fig.~\ref{fig:fig-ferman}.
The main SM backgrounds come from the Drell-Yan, dibosons, $Z$-boson plus jets, and the leptonic top pair 
events. We use \textsc{MadGraph5\_aMC@NLO}~\cite{Madgraph} to generate parton-level events  
and then pass the events to \textsc{Pythia}~\cite{pythia} for showering and hadronization. 
We simulate the detector effects by \textsc{Delphes}~\cite{delphes} and perform the analysis of events 
with \textsc{CheckMATE2}~\cite{checkmate1,checkmate2,checkmate3}. Finally, the significance is obtained 
by
\begin{equation}
Z=\frac{S}{\sqrt{S+B+{(\beta B)}^2}},
\end{equation}
with $S$ ($B$) being the events number of signal (SM background) and $\beta$ being the total systematic uncertainty, taken as $\beta=10\%$ in our calculations.

\begin{figure}[ht]
	\centering
	\includegraphics[width=.45\textwidth]{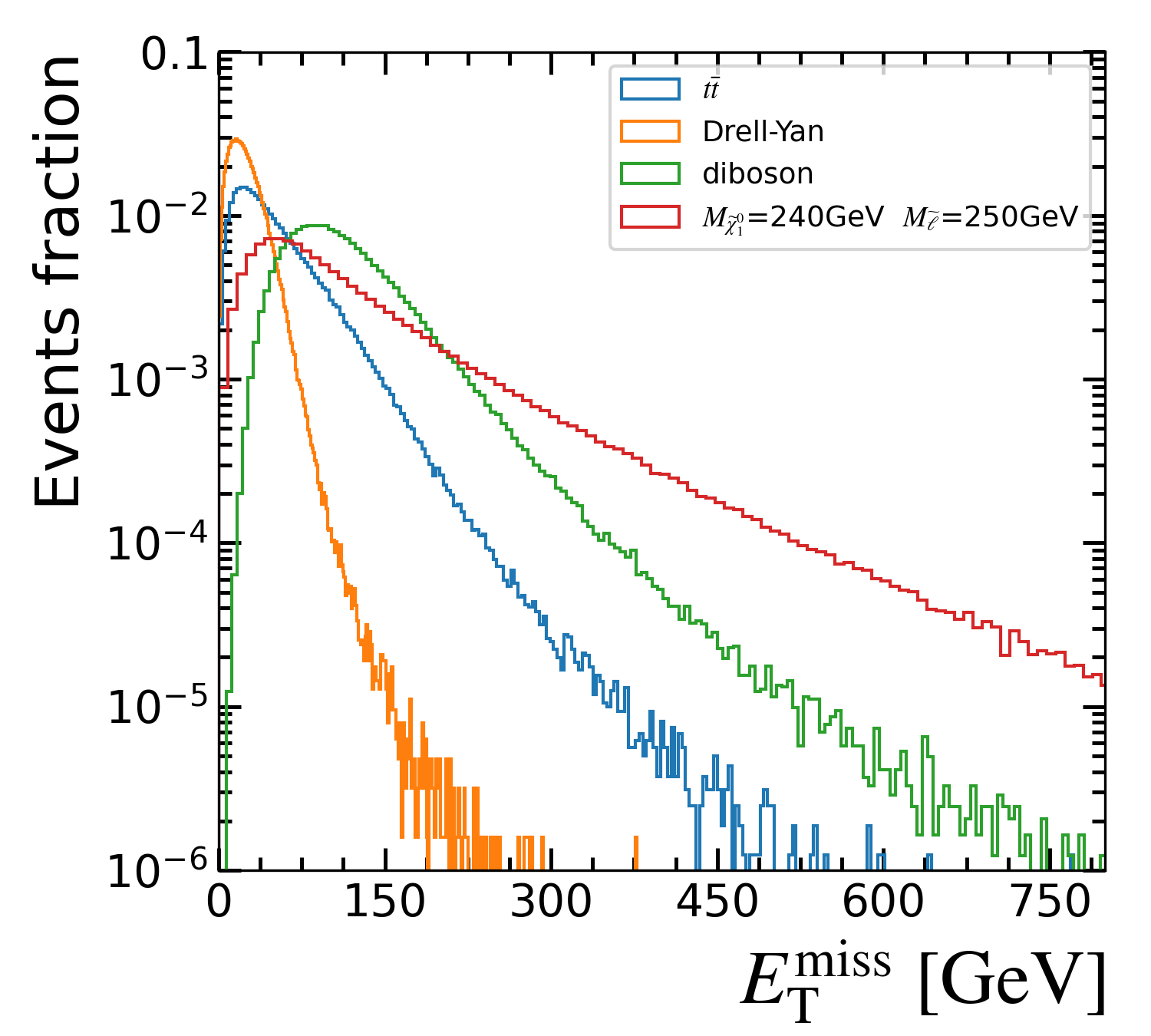}
	\includegraphics[width=.45\textwidth]{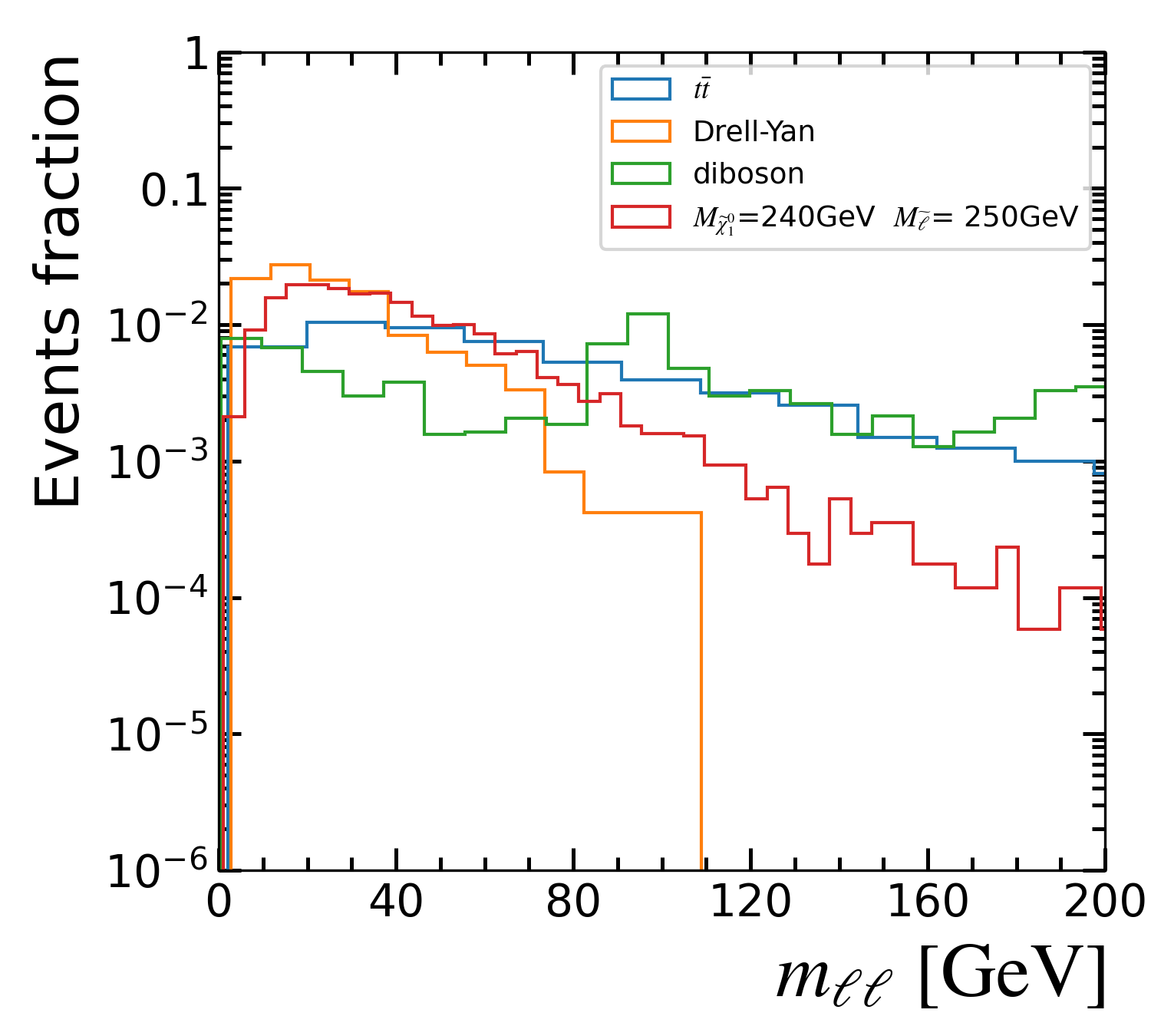}
	\includegraphics[width=.45\textwidth]{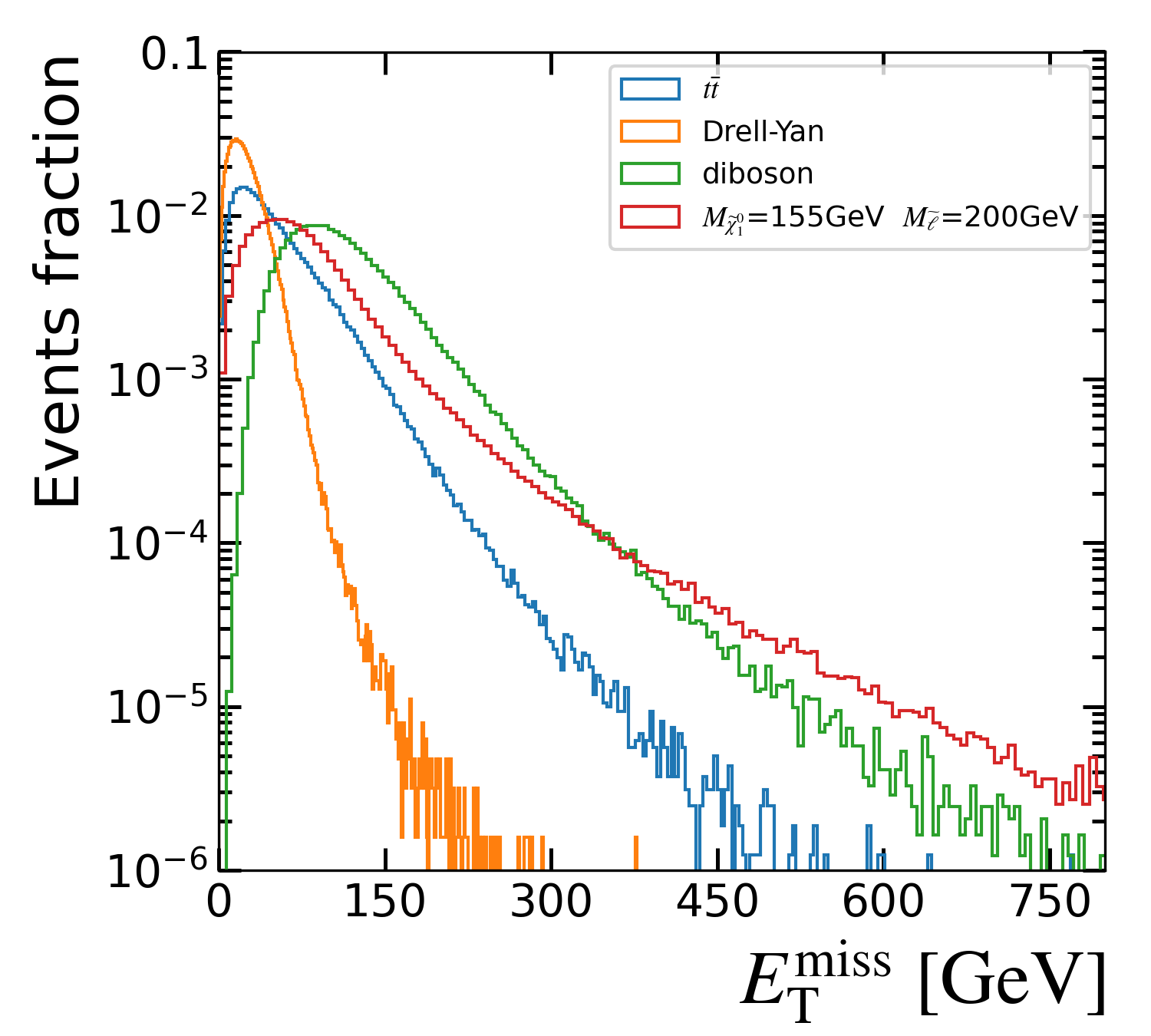}
	\includegraphics[width=.45\textwidth]{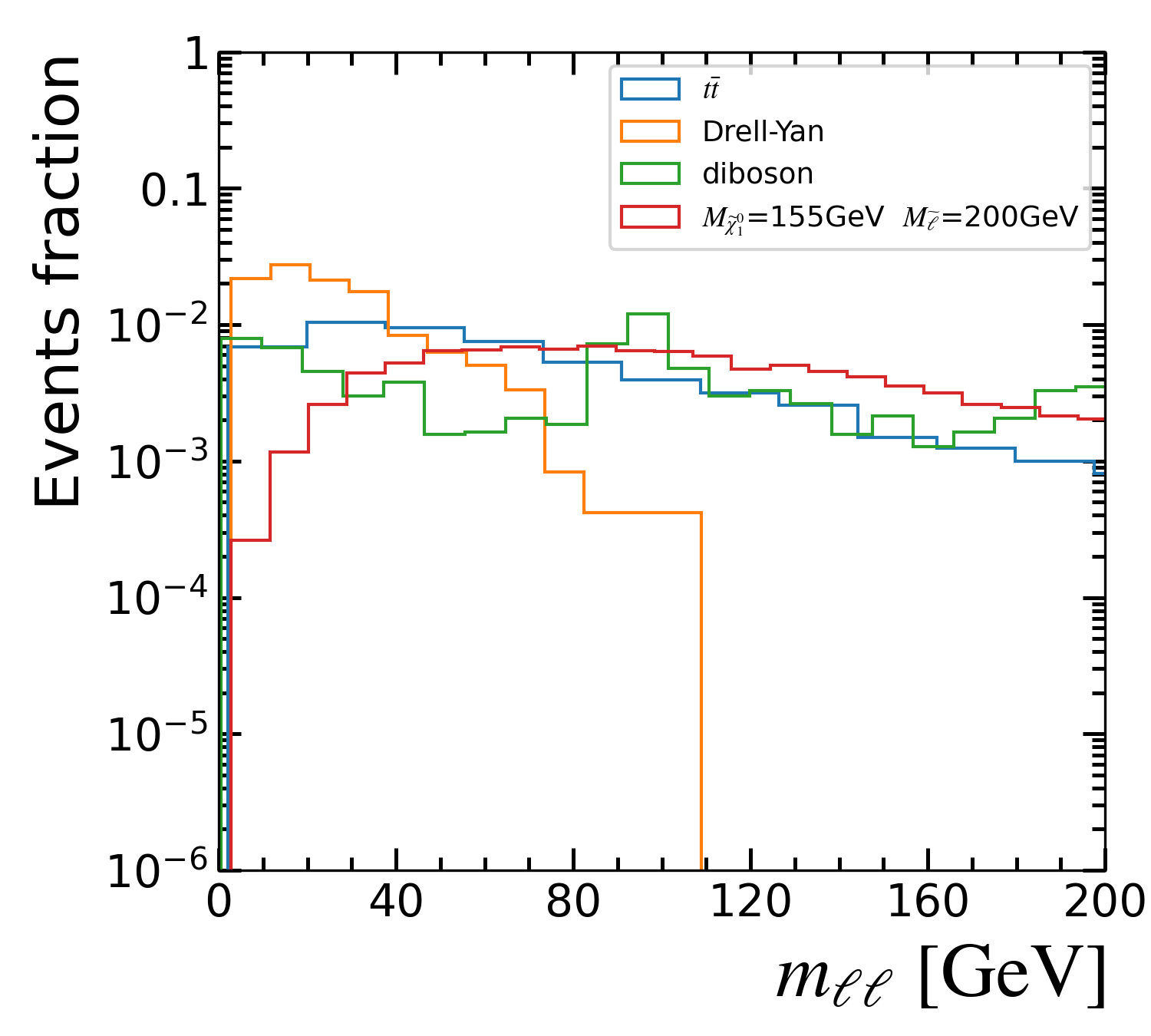}
	\caption{The normalized $ E_{\rm T}^{\rm miss}$ and $m_{\ell \ell}$ distributions of the signal and the 
SM background events at the 14 TeV HL-LHC. The upper and lower panels are for the final states with 
soft and hard 2$\ell$, respectively.}
	\label{fig:events-fraction}
\end{figure}

\begin{figure}[ht]
	\centering
	\includegraphics[width=.7\textwidth]{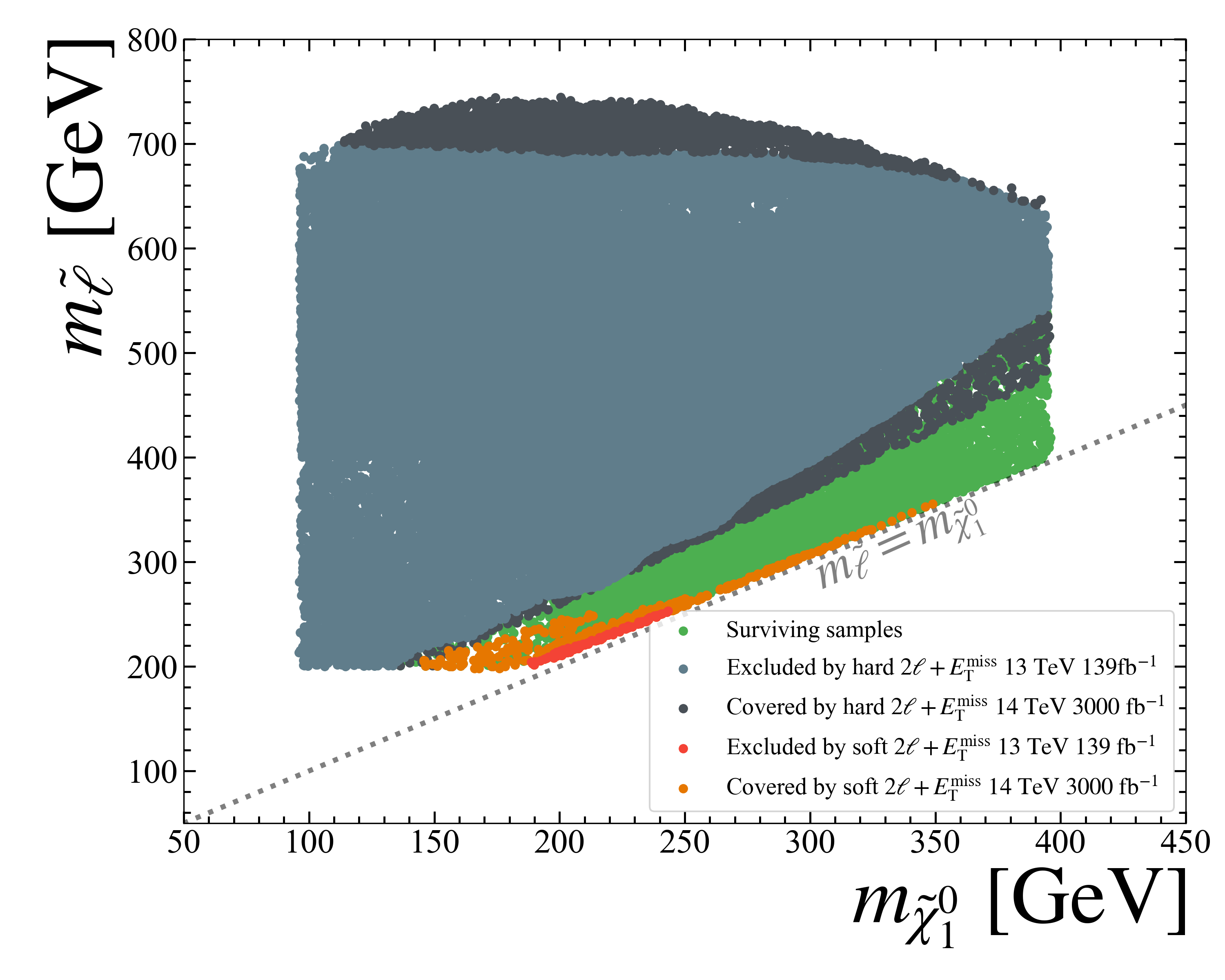}
	\caption{\label{fig-LHC}Same as Fig.~\ref{fig:mass2}, but showing the detection limits of the processes $pp\to \tilde{\ell} \tilde{\ell} + {\rm jets}$ at the HL-LHC. }
\end{figure}

Fig.~\ref{fig:events-fraction} displays the normalized distributions of the missing transverse energy 
and the dilepton invariant mass of the signal and background events. According to the kinematic 
features including those shown in Fig.~\ref{fig:events-fraction}, 
we impose the following event selection criteria. 
\par For the soft dilepton plus missing energy channel, corresponding to the compressed mass spectrum, the event selection is optimized via Recursive Jigsaw Reconstruction (RJR) technique~\cite{Jackson:2016mfb, Jackson:2017gcy} as follows:
\begin{itemize}
    \item[Cut-1:] $E_{\rm T}^{\rm miss}$ trigger. The first cut requires $E_{\rm T}^{\rm miss} > 110~{\rm GeV}$. 
    \item[Cut-2:] Dilepton preselection. Before assigning visible objects to the jigsaw decay tree, the preselection criteria require the signal to have exactly two leptons with $p_{\rm}^{\ell_1} > 5~{\rm GeV}$; the invariant mass $m_{\ell\ell}$ not in the range $[3.0, 3.2]~{\rm GeV}$ to remove contributions from $J/\psi$ decays; $m_{\ell\ell}$ is further required to be smaller than $60~{\rm GeV}$ to suppress the background from on-shell $Z$-boson decays. To reduce background events containing the so-called fake non-prompt leptons, the lepton pairs should be separated: $\Delta R_{\mu\mu} > 0.05$, $\Delta R_{ee} > 0.3$ and $\Delta R_{e\mu} >0.2$. 
    \item[Cut-3:] ISR jet preselection. We require at least one jet with $p_{\rm T}^{j} > 110~{\rm GeV}$ as, for signal events, lepton pairs are boosted by energetic ISR jets. The leading jet is required to be on the different hemisphere from the missing momentum $\Delta \phi(j_1, \vec{p}_{\rm T}^{\rm miss}) > 2.0$, while the additional jets, which are assigned into the ISR system of the compressed decay tree, must satisfy $\Delta \phi(j, \vec{p}_{\rm T}^{\rm miss}) > 0.4$. Furthermore, events with $b$-tagging jets and $p_{\rm T}^{b-{\rm jet}} > 20~{\rm GeV}$ are vetoed. 
    \item[Cut-4:] OSSF. Signal events contain one opposite-sign same-flavor (OSSF) lepton pair plus large missing energy. And to suppress the Drell-Yan backgrounds $Z\to \tau\tau, \tau \to \ell\nu\nu$, the $m_{\tau\tau}$ variable, defined in~\cite{Han:2014kaa, Baer:2014kya, Barr:2015eva}, is required to be negative or to be greater than $160~{\rm GeV}$. 
    \item[Cut-SR:] Signal region. We require $E_{\rm T}^{\rm miss} > 200~{\rm GeV}$. The transverse mass variable $m_{\rm T2}^{m_{\chi}}$~\cite{smt, Lester:1999tx, Barr:2003rg}, with $m_{\chi}$ being an input mass variable,  is used for finer signal selection. First, $m_{\rm T2}^{100}$ is required to be less than $140~{\rm GeV}$ to improve the compressed spectrum, where the chosen value of $m_{\chi} = 100~{\rm GeV}$ is from higgsino mass. Second, the $m_{\rm T2}^{100}$ distribution has an end point of $100~{\rm GeV}$ and the difference $(m_{\rm T2}^{100} - 100)$ reflects the mass splitting between $\tilde{\ell}$ and LSP, which is closely related to the lepton energy. So we require $p_{\rm T}^{\ell_2} > \min\left(20, 2.5+2.5 \times (m^{100}_{\rm T2}-100)\right)$. Finally, the $R_{\rm ISR}$ variable estimated by the RJR technique, which approximately follows the relation $R_{\rm ISR} \sim m_{\tilde{\chi}_1^0} \big/ m_{\tilde{\ell}}$, is required that $\max\left(0.85, 0.98 - 0.02 \times m_{\rm T2}^{100}\right) < R_{\rm ISR} < 1$. 		
\end{itemize}
\par For the hard dilepton channel, the searching strategy is more straightforward. Signal events are required to have two OSSF leptons $\ell_1$ and $\ell_2$ with $p_{\rm T}^{\ell} > 25~{\rm GeV}$ and the invariant mass $m_{\ell\ell} > 121.2~{\rm GeV}$. Any event that contains one $b$-tagging jet with $p_{\rm T}> 20~{\rm GeV}$ is removed to suppress the $t\bar{t}$ background. We also require a large missing transverse energy $E_{\rm T}^{\rm miss} > 110~{\rm GeV}$ and $E_{\rm T}^{\rm miss}\text{-significance} > 10$ ($E_{\rm T}^{\rm miss}$ significance is defined in Eq.~(1) in~\cite{ATLAS-CONF-2018-038}). Finally, $m_{\rm T2} > 100~{\rm GeV}$ is required to reduce SM backgrounds further. 

\begin{table}[h]
\begin{center}
\caption{\label{cutflow}The cut flows for the cross sections in units of fb at the 14 TeV HL-LHC. 
For the soft dilepton signal we choose a benchmark point 
$m_{\tilde\chi^0_1}=240~{\rm GeV}$, $m_{\tilde\ell}=250~{\rm GeV}$, $\tan\beta=50$;
while for the hard dilepton signal we choose a benchmark point $m_{\tilde\chi^0_1}=155~{\rm GeV}$, 
$m_{\tilde\ell}= 200~{\rm GeV}$, $\tan{\beta}=38.5$. }
\resizebox{\textwidth}{!}{
	\begin{tabular}{lp{.1cm}|rp{.05cm}rp{.05cm}rp{.05cm}|r}
		\hline\hline
		\multicolumn{9}{c}{\bf Soft dilepton channel}  \\
	\hline
		\multirow{2}{*}{\bf Cuts} && \multicolumn{5}{c}{\bf SM backgrounds} &&{\bf signal}\\ 
         && $t\bar{t}$ && diboson && Drell-Yan&& (250,240)\\
		\hline
		$E_{\rm T}^{\rm miss}$ Trigger &&138727.9&&5599.09 && 1565.4  &&12.05 \\
		dilepton preselection 
		 &&6041.48  &&361.02  &&104.95  && 4.57\\
		ISR jet preselection 
		&&  195.00 && 40.87 &&47.52  &&2.17 \\
		OSSF
		&& 82.26&&20.08  &&6.93  && 1.96   \\
		signal region
		 && 2.65&& 1.09 && 0.99  && 0.17  \\
		\hline
	    \multicolumn{9}{c}{\bf Hard  dilepton channel}\\
		\hline
		\multirow{2}{*}{\bf Cuts} && \multicolumn{5}{c}{\bf SM backgrounds} &&{\bf signal}\\ 
         && $t\bar{t}$ && diboson && Drell-Yan&& (200,155)\\
		\hline
		$N_{\ell} \geq 2$, $p_{\rm T}^{\ell_1} > 25~{\rm GeV}$, $p_{\rm T}^{\ell_2} > 25~{\rm GeV}$ && 20471.83 && 3622.38 && 16096.39  && 13.92 \\
		OSSF, $N_{b-{\rm jet}}=0$ && 76.28  && 467.62  && 580.97  && 0.57\\
		$m_{\ell\ell}>121.2~{\rm GeV}$ &&  31.17 && 94.06 && 8.90  && 0.32 \\
		$E_{\rm T}^{\rm miss}\text{-significance} > 10$, 
		&& \multirow{2}{*}{4.64} &&\multirow{2}{*}{3.82}  &&\multirow{2}{*}{1.48}  && \multirow{2}{*}{0.06}   \\
        $E_{\rm T}^{\rm miss} > 110~{\rm GeV}$, $m_{\rm T2}>100~{\rm GeV}$ && && && && \\
  \hline\hline
	\end{tabular}
}
\end{center}
\end{table}

\par In Table~\ref{cutflow}, we demonstrate the cut flows for the benchmark points in two channels. 
For both soft and hard dilepton channels, the cut on missing energy $E_{\rm T}^{\rm miss}$ is quite 
crucial to suppress backgrounds. To reduce the huge top pair background, the veto of $b$-jets 
is quite efficient. 
Finally, with all these cuts, we display in Fig.~\ref{fig-LHC} the significance 
of the processes $pp\to j\tilde{\ell}^+\tilde{\ell}^-$ at the HL-LHC. We see that compared with
the current LHC coverage, the HL-LHC can further cover a sizable part of the parameter space
favored by the muon $g-2$ at $2\sigma$ level.  

\par Note that we checked that in this light higgsino scenario the contribution to the $W$-boson mass is 
quite small, much below the magnitude to explain the measured value by CDF II~\cite{CDF:2022hxs}.
The reason is that, as found in \cite{Yang:2022gvz}, the SUSY contribution to the $W$-boson mass mainly comes from
the stops and the explanation of  CDF II result needs a stop around 1 TeV. In our scenario, stops are assumed to be quite heavy.
Also, this light higgsino scenario cannot jointly explain the electron and muon $g-2$ anomalies (with the fine structure constant from the Berkeley experiment~\cite{Parker:2018vye}, the SM value of electron $g-2$ is above the experimental value~\cite{Hanneke:2008tm} by $2.4\sigma$).  As shown in 
~\cite{Li:2021koa,Li:2022zap,Dutta:2018fge,Endo:2019bcj,Ali:2021kxa,Yang:2020bmh,Cao:2021lmj}, a joint explanation needs a rather special parameter space in SUSY. 

\section{\label{sec:conclusions}Conclusions}
We examined the light higgsino scenario in light of the muon $g-2$ data. 
The dark matter constraints on the light higgsino-like LSP were also taken into account. 
Assuming a light higgsino mass parameter $\mu$ in the range of 100-400 GeV while keeping gaugino
mass parameters above TeV, we explored the parameter space.
We found that, to explain the muon $g-2$ anomaly at $2\sigma$ level, the winos and sleptons are 
respectively upper bounded by 3 TeV and 800 GeV. Then, for the  light higgsino-like LSP, 
we found that it can sizably scatter with 
nucleon and thus the allowed parameter space can be covered almost fully by the future LZ dark matter detection 
project. Finally, for the light leptons we performed a Monte Carlo simulation for their pair production
at the HL-LHC and found that compared with the current LHC limits, the HL-LHC can further cover a sizable part 
of the parameter space. 

\addcontentsline{toc}{section}{Acknowledgments}
\acknowledgments
This work was supported by the National Natural Science Foundation of China under Grant No. 12275066 and 
by the startup research funds of Henan University. 
 
\bibliography{references.bib}
\bibliographystyle{CitationStyle}

\end{document}